\documentclass[conference]{IEEEtran}
\IEEEoverridecommandlockouts
\usepackage{booktabs}
\usepackage{cite}
\usepackage{amsmath,amssymb,amsfonts}
\usepackage{graphicx}
\usepackage{textcomp}
\usepackage{xcolor}
\usepackage{multirow}
\usepackage{xurl}
\usepackage{enumitem}
\usepackage{algorithm}
\usepackage[noend]{algpseudocode}
\usepackage{bbm}
\usepackage{subcaption}
\usepackage{amsthm}
\usepackage{wrapfig}
 
\begin{document}

\title{zk-IoT: Securing the Internet of Things with Zero-Knowledge Proofs on Blockchain Platforms}

\author{\IEEEauthorblockN{Gholamreza Ramezan}
\IEEEauthorblockA{\textit{FidesInnova} \\
g2n@fidesinnova.io}
\and
\IEEEauthorblockN{Ehsan Meamari}
\IEEEauthorblockA{\textit{FidesInnova} \\
e2i@fidesinnova.io}
}
\maketitle

\begin{abstract}
This paper introduces the zk-IoT framework, a novel approach to enhancing the security of Internet of Things (IoT) ecosystems through the use of Zero-Knowledge Proofs (ZKPs) on blockchain platforms. Our framework ensures the integrity of firmware execution and data processing in potentially compromised IoT devices. By leveraging the concept of ZKP, we establish a trust layer that facilitates secure, autonomous communication between IoT devices in environments where devices may not inherently trust each other. The framework includes zk-Devices, which utilize functional commitment to generate proofs for executed programs, and service contracts for encoding interaction logic among devices. It also utilizes a blockchain layer and a relayer as a ZKP storage and data communication protocol, respectively. Our experiments demonstrate that proof generation, reading, and verification take approximately 694, 5078, and 19 milliseconds in our system setup, respectively. These timings meet the practical requirements for IoT device communication, demonstrating the feasibility and efficiency of our solution. The zk-IoT framework represents a significant advancement in the realm of IoT security, paving the way for reliable and scalable IoT networks across various applications, such as smart city infrastructures, healthcare systems, and industrial automation.
\end{abstract}


\section{Introduction}\label{sec:introduction}
\par Internet of Things (IoT) refers to a network of interconnected physical devices, objects, or ``things" that are embedded with sensors, software, and network connectivity, allowing them to collect and exchange data over the Internet. IoT applications range from items such as wearable devices, smart home appliances, and drones to more specialized industrial equipment and vehicles. IoT technology enables these devices to communicate with each other and with central systems, enabling data collection, monitoring, and automation. 

\par A significant challenge facing IoT is ensuring trust. Without robust security mechanisms to ensure mutual trust among IoT devices, the functionality of each IoT device remains isolated, limiting the full potential that could be realized through integrated IoT solutions. An adversary may interfere with the device's firmware or modify its hardware or software to generate incorrect output from a legitimate device, posing a threat to the integrity of the device's computations. Additionally, the adversary may introduce a rogue IoT device to generate and transmit data to other users or IoT actuators, compromising the system's integrity. Hence, the concern about IoT \textit{functions}, which is a result of executing its firmware or program, hinders the ability to fully leverage the interconnected capabilities of IoT device \cite{JCST-2312-14035}.

\par Addressing the above-mentioned challenge not only ensures the integrity of executed computations but also paves the way for autonomous IoT device communication. The scheme enables mutually distrustful IoT devices, that are from different manufacturers and are in separate IoT ecosystems such as Ring and Ecobee IoT systems, to perform distributed computations that run indefinitely, while ensuring that the correctness of every intermediate state of the computation can be verified efficiently. For instance, in the electric vehicle sector, represented by companies like Tesla, a decentralized trusting IoT framework would enable cars to securely communicate and receive real-time information about road conditions from vehicles traveling in the opposite direction. Additionally, it ensures that various components within a vehicle can authenticate whether commands received from the car's main computer are the result of legitimate computation (not hacked system), thereby establishing trust in the system's integrity.

\par The mentioned trust challenge can be addressed through the use of IoT devices built within a single IoT device manufacturer ecosystem. However, IoT devices are engineered across diverse ecosystems by various manufacturers, leading to a situation where the specific functionalities encoded within their firmware are not consistently disclosed. However, the principal features of their operational capabilities are generally recognized. Furthermore, companies may be reluctant to share detailed information about the architecture or firmware of their devices.

\par
\begin{figure}[tbp]
\centering
\includegraphics[width=0.40\textwidth]{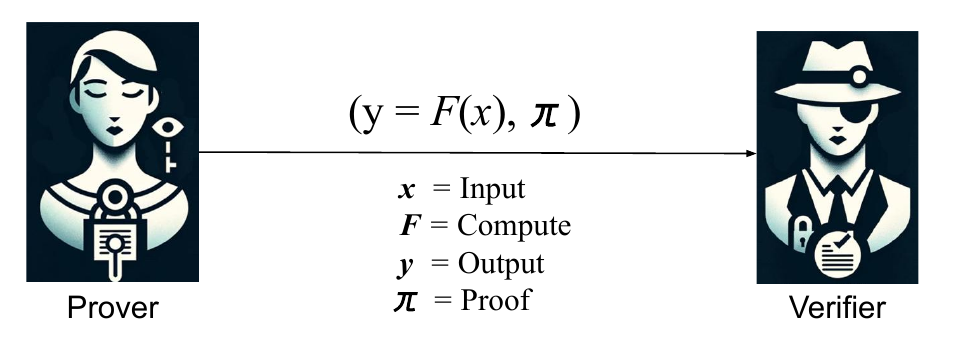}
\caption{Key components of zero-knowledge proofs (ZKPs).}
\label{ZKProles}
\end{figure}

 Zero-Knowledge Proof (ZKP) is a  cryptographic technique enabling one party (the prover) to demonstrate the truth of a statement to another (the verifier) without revealing any information beyond the statement's truth itself. Fig. \ref{ZKProles} illustrates the roles of the prover and verifier in a ZKP scenario. Here, the prover shares only the output of a specific function, without disclosing the input (secret), along with a proof. In the ZKP realm, \textit{functional} commitment is a protocol that hides a function $f$ from the verifier, yet allows the verifier to confirm the correctness of the function's execution without needing to rerun it or know the precise function \cite{de2023functional, cryptoeprint:2021/1342}.

\vspace{-3pt}

\par Tackling the issue of computational integrity within the IoT realm paves the way for the development of solutions that facilitate \textit{autonomous device communication} among devices that may not inherently trust each other. Moreover, various traditional IoT attacks can be addressed such as the injection of malicious code that can alter processing results. 

\par In the realm of automatic IoT device communication, every device at every step of a communication process should trust the data generated by the previous step's IoT device. proof-carrying data (PCD) primitive enables the composition of proofs to attest to the correctness of computations \cite{Chiesa2010}. This primitive is particularly significant in scenarios where data and computations need to be verified across multiple parties (e.g., IoT devices) that may not fully trust each other. PCD allows for constructing proofs that not only certify the correctness of a computation but also carry data relevant to that computation. These proofs can be composed, meaning that the output of one computation, along with its proof, can be used as an input to another computation, creating a chain of trustworthy computations. This concept is crucial in decentralized systems, such as blockchain technology, where trust and verification are distributed among many participants \cite{Chiesa2010}. 

\par
\begin{figure}[tbp]
\centering
\includegraphics[width=0.50\textwidth]{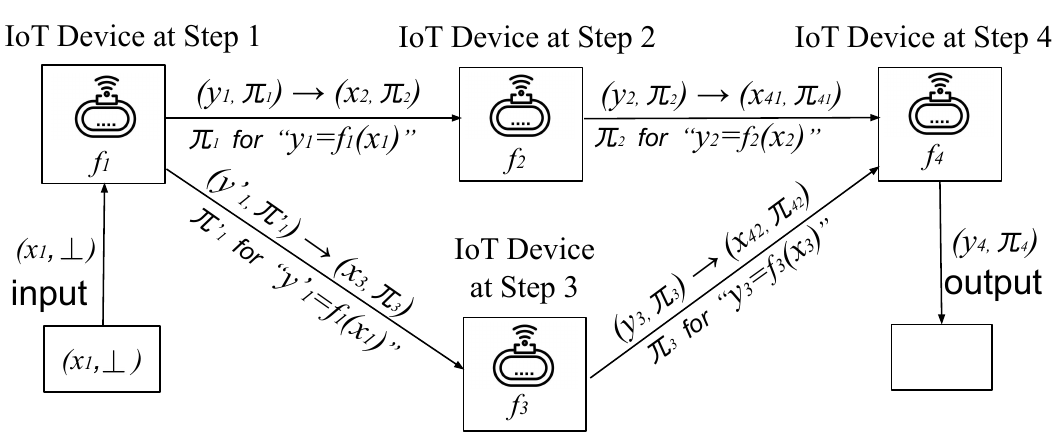}
\caption{A sample Internet of Things (IoT) devices setup.}
\label{PCD}
\end{figure}

 Fig. \ref{PCD} demonstrates the application of PCD in the IoT domain. The process initiates with a known input $x_1$. This input is then processed by an IoT device, which applies function $f_1$ to $x_1$ along with any other internal inputs it may receive from its sensors. Subsequently, the device produces an output and a proof verifying that the function has been executed correctly. This sequence persists until the final IoT device in the chain generates the output and proof for its operation.

\vspace{-3pt}

\par This paper proposes a framework designed to automate communication among untrusted IoT devices, thereby expanding IoT functionalities and enabling innovative IoT services that are not achievable by an isolated IoT device or a collection of IoT devices controlled by a central authority. The main contributions of this paper are as follows. 
\begin{itemize}
\item We introduce the zk-IoT framework, designed for automatic communication between untrusted IoT devices. It comprises components including zk-Devices, blockchain nodes, relayers, and service contracts, as well as a blockchain protocol and a ZKP scheme.
\item We introduce a ZKP scheme designed to ensure the integrity of IoT firmware execution through functional commitment (FC) within a distributed IoT setup.
\item We also introduce several key concepts: a service contract, which encodes the interaction logic among IoT devices; zk-Devices, which utilize ZKP to generate proofs of IoT firmware execution; and a blockchain protocol, designed for automatic communication between untrusted IoT devices.
\item We include a performance evaluation within a blockchain context, demonstrating the feasibility of implementing the zk-IoT framework. 
\end{itemize}
This paper is structured as follows. In Section 2, we review the related works.  Section 3 will introduce the proposed zk-IoT framework. In Section 4, we perform a performance analysis. Section 5 concludes.

\section{Related work}
This section provides an overview of existing ideas on how ZKP can empower IoT and introduces our proposal to further enhance IoT using ZKP.

Traditional authentication methods use robust encryption protocols to secure data transfer, ensuring protection from unauthorized access and interception. However, ZKPs revolutionize this approach. Chen et al. \cite{Chen2023-A-Survey-on-ZK-Authentication-for-IoT} review ZKP-based IoT authentication, highlighting its benefits over traditional methods in terms of anonymity and reduced computational complexity. Walsh et al. \cite{walshe2019non}, proposed non-interactive ZKPs in a Merkle tree structure for authentication, along with methods combining ring signatures with ZKPs for enhanced privacy \cite{mahmood2020privacy}. W{\"o}hnert et al. \cite{wohnert2020secure} proposed ZKP-based algorithms for verifiable identity in IoT, emphasizing accountability and integrity. Access control mechanisms in IoT using ZKPs have also been explored, leveraging Ethereum smart contracts and the Groth16 proof generation mechanism \cite{song2021access}. Verifiable anonymous identity management (VAIM) models have been proposed, focusing on privacy-preserving user-centric identity verification \cite{ra2021vaim}.

\par In the realm of blockchain-based IoT, challenges related to data immutability and General Data Protection Regulation (GDPR) compliance have prompted research into blockchain erasure models, combining ring signatures and ZKPs for enhanced privacy \cite{elgabri2023blockchain, rani2023multi}. Furthermore, balancing privacy with accountability in blockchain identity management is essential, as asserted by studies advocating for traceability services in ZKP-based systems \cite{2021Balancing-Privacy-and-Accountability-in-Blockchain-Identity-Management, Luong2023-Privacy-Preserving-Identity-Management-System-on-Blockchain-Using-Zk-SNARK}. These systems combine authentication with privacy and traceability, using shared secrets among verification nodes and central authority oversight to detect and manage malicious behavior.

\par ZKPs can also be instrumental in proving the validity of the executed process without revealing the actual software code or detailed information about the process. This method improves security as it eliminates the risk of exposing software integrity mechanisms that hackers could potentially exploit to circumvent the attestation process. In \cite{kang2022zk}, a novel approach for attested cameras is proposed to verify that images were taken by a camera. Zero-knowledge succinct non-interactive argument of knowledge (zk-SNARK) allows verification of image transformations non-interactively with only standard cryptographic hardness assumptions. Unfortunately, this work does not preserve input privacy, is impractically slow, and/or requires custom cryptographic arguments \cite{fang2023application}.

In addition to its previous applications, ZKP can be utilized to enable IoT devices to demonstrate that they have executed a particular function (firmware) on specific data, thereby ensuring the integrity of both the data and the function. This application, when combined with blockchain technology, enhances the capabilities of IoT for future applications. The focus of this paper is to propose an architecture that facilitates this enhancement.

\section{zk-IoT Framework}
\begin{figure}[tbp]
\centering
\includegraphics[width=0.45\textwidth]{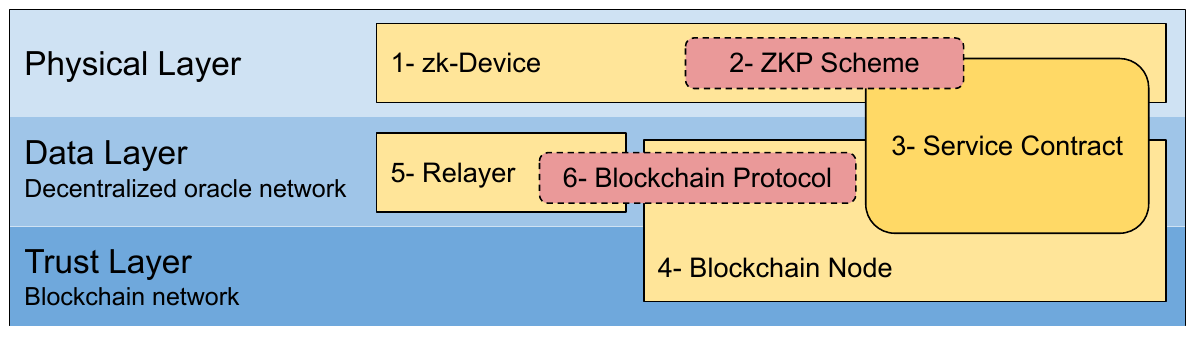}
\caption{The main components of the zk-IoT framework.}
\label{zkIoTFramework}
\end{figure}

The Zero-Knowledge Internet of Things (zk-IoT) framework ensures the integrity of firmware execution in IoT devices and facilitates automatic communication between untrusted IoT devices. Hence, it enables IoT devices to collaboratively offer innovative services to users by integrating functionalities and processed data across all devices. As it is shown in Fig. \ref{zkIoTFramework}, the zk-IoT framework consists of six main elements: zk-Devices, ZKP scheme, service contracts,  blockchain nodes, relayers, and blockchain protocol. The architecture is structured across three tiers. The top tier, known as the physical layer, houses the zk-Devices. The middle tier, or the data layer, includes the relayers from a decentralized oracle network, which facilitates data transfer and bridges the physical world with the blockchain realm. The bottom tier, the blockchain layer, serves as the trust foundation for the framework. It incorporates blockchain nodes to store ZKPs and uses smart contracts to facilitate the data transfer process.

\vspace{-3pt}

\subsection{zk-Device}
\begin{figure}[h]
\centering
\includegraphics[width=0.4\textwidth]{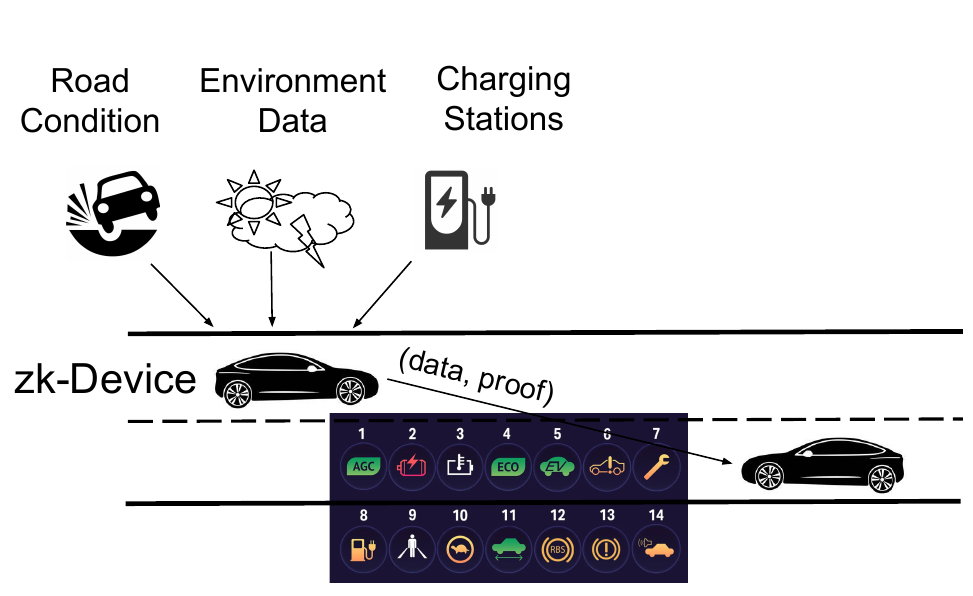}
\caption{zk-Device}
\label{zk-Device-Tesla}
\end{figure}

In the zk-IoT framework, every IoT device is equipped with a ZKP generator to validate the integrity of its executed programs. Fig. \ref{zk-Device-Tesla}, shows the inputs for the computation originating from IoT sensors, while the program is stored within the IoT device's memory.

\par The zk-Device ensures that its output originates from a circuit designed by the manufacturer, certifying that the IoT device operates exactly as intended by its maker. It also guarantees that both the program and its inputs remain unchanged. As a result, the device's output is reliable and trustworthy.  Upon completion of the computation, the output, along with the generated proof, is transmitted from the IoT device to the external physical world. It is important to note that the proof guarantees the execution of the program and the data inputted into the device either after analog-to-digital converters (ADC) or before digital-to-analog converters (DAC). This means that the boundary of the computation is confined to the digital part of the system where computations can be performed.

\vspace{-3pt}

\subsection{ZKP Scheme}
\par zk-Devices implement a ZKP scheme that utilizes FC and PCD to verify a chain of computations across a series of IoT devices interacting during an automation process. As discussed in section \ref{sec:introduction}, the scheme enables mutually distrustful IoT devices to perform distributed computations that run indefinitely while ensuring that the correctness of every intermediate state of the computation can be verified efficiently. In other words, the scheme is designed to verify a claim of the form $f^T(x_{11},.., x_{1m})=y_T$ where $m$ is the number of IoT devices and $f^T$ denotes a computation of an automation process with $T$ steps.  Also, $(x_{11},.., x_{1m})$ is the input to the first IoT device in the automation process. $y_T$ is the final process result. Without loss of the generality, the scheme is described on one IoT device at Step $i$ of the computation with a private function $f_i$ and public inputs $(x_{i1},...,x_{im})$  as shown in Fig. \ref{PCDScheme}. This scheme is structured as a tuple (\textsc{Setup}, \textsc{Commit}, \textsc{Proof}) where: 

\begin{figure}[tbp]
\centering
\includegraphics[width=0.35\textwidth]{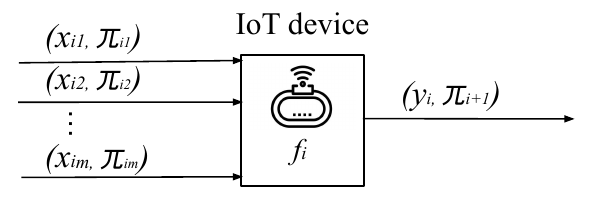}
\caption{A sample IoT device at step $i$ in an automation process.}
\label{PCDScheme}
\end{figure}
\vspace{-3pt}

\begin{itemize}
\item \textsc{Setup}($1^\lambda, C_{f_i}) \rightarrow (pp_i, pk_i, vk_i)$: for a security parameter $\lambda$ and circuit $C_{f_i}:= f_i(x_{i1},..., x_{im})-y_i$, produces the public parameter $pp_i$, proving key $pk_i$ and verification key $vk_i$.
\item \textsc{Commit}($pp_i, A_i, B_i, C_i) \rightarrow (Com_{A_i}, Com_{B_i}, Com_{C_i})$: Prover represents circuit $C_{f_i}$ by matrices $A_i, B_i, C_i$ in $\mathbb{F}^{n\times n}$ where $\mathbb{F}$ is a field that wires of circuit carry their values of it and  $n$ is the number of constraints such that $C_{f_i}(x_{i1},..., x_{im}, y_i)=0$ iff there is $(w_{i1},..., w_{in-m})$ in $\mathbb{F}^{n-m}$ such that $(A_iz_i)o(B_iz_i)=C_iz_i$ where $z_i=(x_{i1},..,x_{im}, w_{i1},..., w_{in-m})$ and operation $o$ denotes to component-wise product of two equal size vectors. This function produces commitment $Com_{A_i}=(Com_{Row_{A_i}}, Com_{Col_{A_i}}, Com_{Val_{A_i}})$ where: 
\begin{itemize}
    \item $Com_{Row_{A_i}}$ is a  commitment to polynomial $Row_{A_i}$ that is made by non-zero elements in $A_i$ such that $Row_{A_i}(\gamma ^j)=\omega ^{r_j}$ where $r_j$ is the row number of $j^{th}$ non-zero element and $\gamma$ is a generator of a multiplicative subgroup of field $\mathbb{F}$ of order $m$. Also, $m$ is the maximum number of non-zero elements of matrices $A_i$, $B_i$ and $C_i$. Also, $\omega$ is a generator of a multiplicative subgroup of field $\mathbb{F}$ of order $n$.
    \item $Com_{Col_{A_i}}$ is a commitment to polynomial $Col_{A_i}$ that is made by non-zero elements in $A_i$ such that $Col_{A_i}(\gamma ^j)=\omega ^{c_j}$  where $c_j$ is the column number of the $j^{th}$ non-zero element. 
    \item $Com_{Val_{A_i}}$ is a commitment to polynomial $Val_{A_i}$ that is made by non-zero elements in $A_i$ such that $Val_{A_i}(\gamma ^j)=v_j$ where $v_j$ is the value of $j^{th}$ non-zero element. The $Com_{B_i}$ and $Com_{C_i}$ are also made similarly.
\end{itemize}
\item \textsc{Proof}($pk_i, A_i, B_i, C_i) \rightarrow (\pi_{A_i}, \pi_{B_i}, \pi_{C_i})$:
produce $\pi_{A_i}$ and $\pi_{B_i}$ to prove $A_i$ and $B_i$ are lower triangular matrix based on protocol "SLT Test" in \cite{cryptoeprint:2021/1342} and $\pi_{C_i}$ to prove $C_i$ is a diagonal matrix based on protocol "Diag Test" in \cite{cryptoeprint:2021/1342}.\\ 
\item \textsc{Verify}($vk_i, \pi_{A_i}, \pi_{B_i}, \pi_{C_i}) \rightarrow \{ accept, reject\}$: Verifies that matrices $A_i$ and $B_i$ is lower triangular matrix based on protocol "SLT Test" in \cite{cryptoeprint:2021/1342} and  $C_i$ is a diagonal matrix based on protocol "Diag Test" in \cite{cryptoeprint:2021/1342}. 
\end{itemize}

\subsection{Service Contract}
Realizing automated IoT device communication requires a mechanism to encode the interaction logic among IoT devices. This is achieved through what we call a Service Contract. A Service Contract specifies the protocol for IoT device communication, detailing how data produced by one IoT device can be utilized by others. Additionally, it ensures the integrity of the data generated at each step by verifying the ZKPs produced by IoT devices engaged in a contract. Service contracts unlock the potential of IoT devices by amalgamating the functions of individual, isolated devices to create new, integrated functionalities. For example, employing a service contract to merge the functionalities of temperature sensors in a wide residential area with a siren in a fire department can create an effective fire detection system. Similarly, integrating a temperature display in a hydro company could lead to an efficient energy management system. These types of integrations serve as prime examples of how service contracts can innovatively augment the capabilities of IoT devices.
\par Algorithm \ref{alg:servicecontract} outlines the basic logic contained in a service contract, which processes a set of inputs including GPS coordinates, a timestamp, a road status (possible car-accident or road problem), the device type, and a proof verifying the integrity of the firmware execution that produced the input data. The process begins by verifying the proof. If the device is identified as a Tesla and the data's location is pinpointed to Seattle, the algorithm then forwards the road condition information and timestamp to the next device in the service chain. The practical implementation of this process for a multi-sensor system will be explored in the performance evaluation section, with examples using JavaScript programming.
\begin{algorithm}
\caption{Service Contract Pseudocode.}\label{alg:servicecontract}
\textbf{Input:} Proof $P$\\
\textbf{Input:} GPS $L$\\
\textbf{Input:} Timestamp $T_i$\\
\textbf{Input:} Collision $C_i$\\
\textbf{Input:} Device Type $D$\\
\textbf{Output:} Temperature $T_o$\\
\textbf{Output:} Temperature $C_o$
\vspace{-0.2em}
\begin{algorithmic}[1]
\Require $Verify(P)==valid$
\Require $Compliance(P,D,'Tesla')==valid$
\If{city($L$)=='Seattle' and   type($D$)=='Tesla'}
    \State Return  $T_i$ as $T_o$
    \State Return  $C_i$ as $C_o$
\EndIf
\end{algorithmic}
\end{algorithm}
\vspace{-0.2em}

\subsection{Blockchain node}
The blockchain node is a critical component in blockchain technology, responsible for executing blockchain algorithms, including consensus algorithms, smart contract execution, block creation, and transaction verification. It also maintains the blockchain data in what we call 'hot storage,' which is replicated across all nodes in the network.

\par Within the IoT domain, each device is required to transmit its physical data to a storage system and may receive commands from it. We call it 'cold storage'. This data exchange is secure and exclusive, establishing a direct trust relationship between a node and an IoT device.  

\par In the zk-IoT framework, IoT functionalities are integrated into the blockchain node, creating a node that features two distinct databases for hot and cold storage. Hot storage is dedicated to blockchain data, such as ZKPs, transactions, and smart contracts. Cold storage, on the other hand, houses a local copy of data from zk-Devices and service contracts. Unlike hot storage, cold storage is not automatically shared across the network. Instead, data transfer between source and target nodes occurs through a relayer in a decentralized oracle network, activated by a service contract.

\subsection{Relayer}
Decentralized oracle networks (e.g., Chainlink) act as essential connectors between blockchain systems and the external environment, enabling the seamless flow of data from off-chain sources to on-chain smart contracts. This process allows blockchain applications to interact with real-world data. Within these networks, relayers are key components, facilitating communication among nodes. 

\par The zk-IoT framework utilizes a relayer to facilitate communication among blockchain nodes, streamlining the process and ensuring integrity and trust.

\subsection{Blockchain protocol}
In the context of zk-IoT, every automated interaction involves a series of steps designed to facilitate communication and execution among different IoT devices. These interactions are grounded in the fundamental premise of a straightforward exchange between two IoT devices. Without loss of generality, this section will focus on the interaction of only two devices under a service contract. The first device generates data, which the second device then receives, verifies, and uses as input for its own firmware execution, subsequently generating new data along with a ZKP for the next device.
\par
\begin{figure}[ht]
\centering
\includegraphics[width=0.5\textwidth]{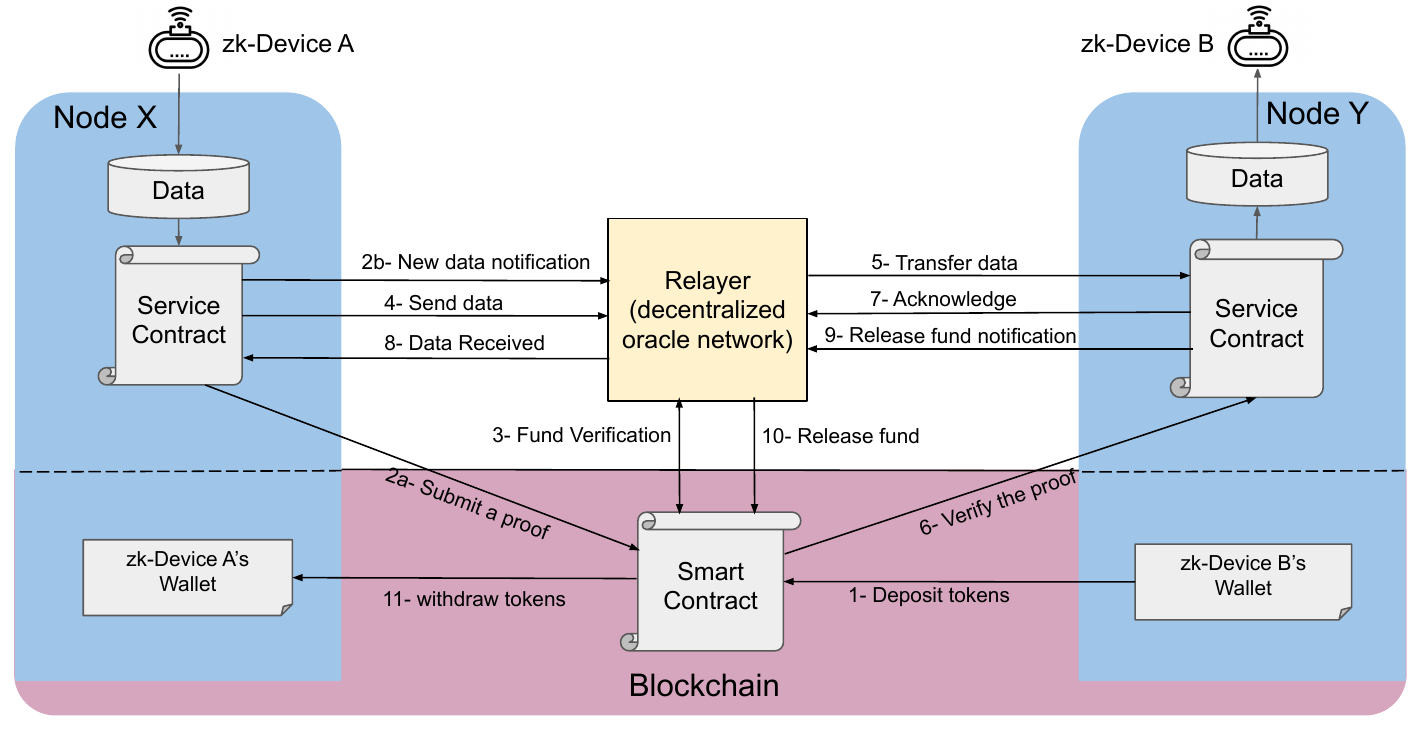}
\caption{zk-IoT communication on blockchain layers.}
\label{zkIoTBlockchain}
\end{figure}

 The workflow of the blockchain protocol is detailed in Figure \ref{zkIoTBlockchain}. The process is initiated by a service contract but requires sufficient funding from the second device, which receives the data. To illustrate this, we describe the process beginning with zk-Device B depositing funds into a smart contract distinct from the service contract. This dedicated smart contract serves solely as a reserve pool specifically for data transactions and is designed to hold and disburse funds according to a predefined sequence. 
\begin{enumerate}
\item \textbf{Deposit Tokens:} The receiving IoT device (zk-Device B) deposits sufficient funds into a smart contract for data transfer.
\item \textbf{Submit Proof and New Data Notification:} Upon transmitting new data, zk-Device A activates a function within Blockchain Node X's service contract to prompt a signal to the Relayer and submits proof to the smart contract.
\item \textbf{Fund Verification:} The Relayer checks if the funds for the transaction are adequate and confirms sufficient funds.
\item \textbf{Transfer Data:} Service Contract X relays the data from zk-Device A to the Relayer.
\item \textbf{Pass Data:} The Relayer passes the data to Service Contract Y on Blockchain Node Y.
\item \textbf{Verify Proof:} Service Contract Y checks the proof.
\item \textbf{Acknowledgment:} Service Contract Y sends an acknowledgment to the Relayer.
\item \textbf{Data Received Notification:} The Relayer sends a notification to Service Contract X.
\item \textbf{Release Fund Notification:} Following successful data processing, Service Contract Y instructs the Relayer to issue a Fund Release command to the Smart Contract.
\item \textbf{Release Funds:} The Relayer requests the release of funds.
\item \textbf{Withdraw Tokens:} zk-Device A can withdraw the funds from the smart contract.
\end{enumerate}

\vspace{-3pt}

\section{Performance Evaluation}
We have developed a preliminary version of the zk-IoT system as described below. The service contracts were crafted using JavaScript due to available ZKP libraries like SnarkJS that enable the users to verify the proof while letting the user to encode the complex interactions required for automating processes across multiple IoT devices. The underlying blockchain network utilizes Ethereum technology, supplemented by a specially modified blockchain node designed to run IsolatedVM, enabling it to execute JavaScript-based service contracts. Also, the blockchain node supports the Message Queuing Telemetry Transport (MQTT) protocol for the non-blockchain part to communicate with zk-Devices. The desired framework utilizes a decentralized oracle network for data transfer, however, the preliminary version incorporates relayer functionality inside our blockchain node. Additionally, the zk-Device is built around the ESP32 micro-controller, Espressif, integrating ZKP libraries. 

\par
\begin{figure}[h]
\centering
\includegraphics[width=0.5\textwidth]{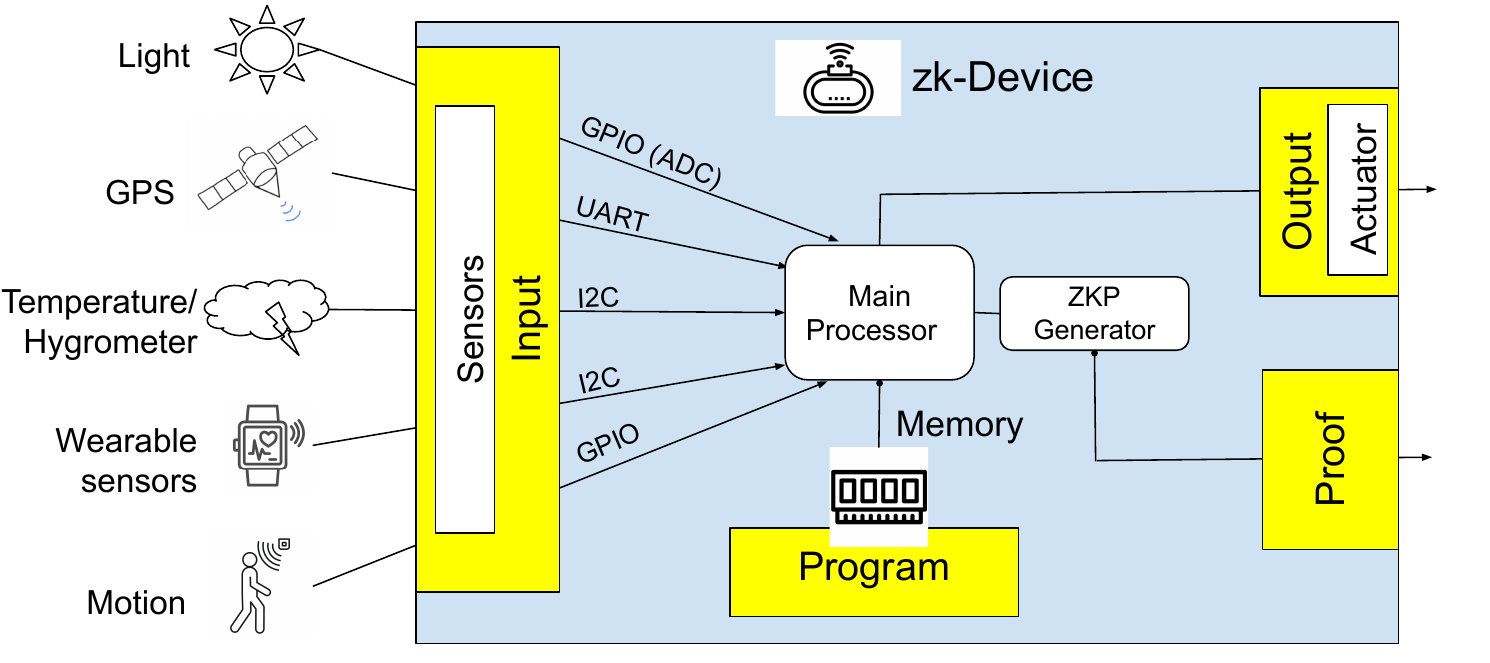}
\caption{ZKP-enabled environmental sensor: general architecture.}
\label{zk-Device}
\end{figure}

\begin{figure}[h]
\centering
\includegraphics[width=0.3\textwidth]{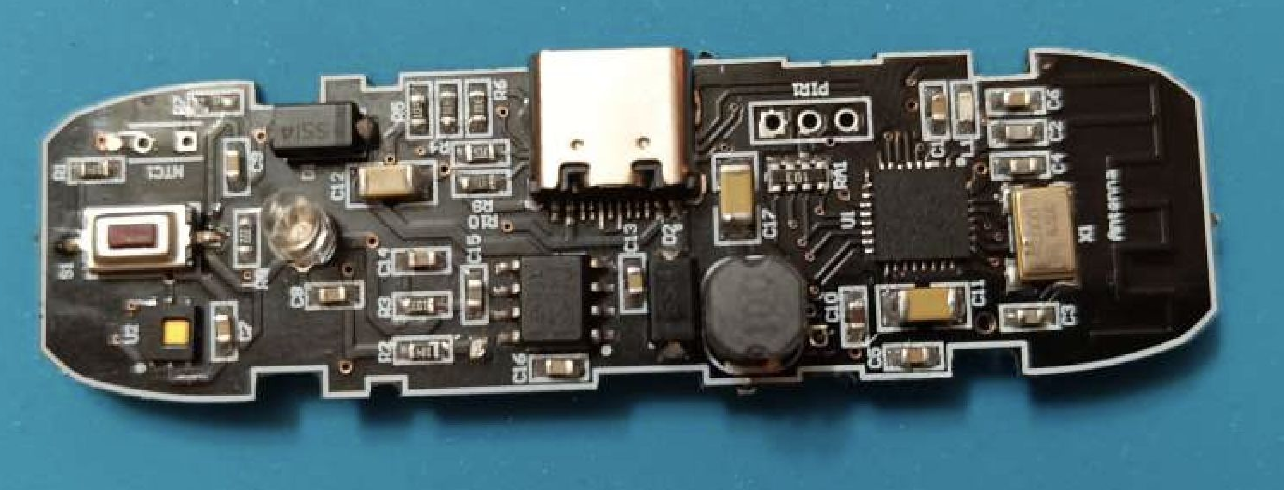}
\caption{ZKP-enabled environmental sensor.}
\label{multisensor1}
\end{figure}

 Figures \ref{zk-Device} and \ref{multisensor1} showcase the schematic and photograph, respectively, of a ZKP-enabled environmental sensor hardware design. This design includes functionalities such as a temperature sensor, hygrometer, motion detector, and smart button. For the purposes of this performance evaluation report, the demonstration of ZKP functionality is confined to the smart button component, which generates a notification upon being pressed. The system has been developed using two ZKP methods: the halo2\_proofs library (version 0.3.0) in Rust, and Circom. We implemented a straightforward communication protocol between two IoT devices.  

\vspace{-3pt}

\vspace{-3pt}

\par
\begin{figure}[h]
\centering
\includegraphics[width=0.47\textwidth]{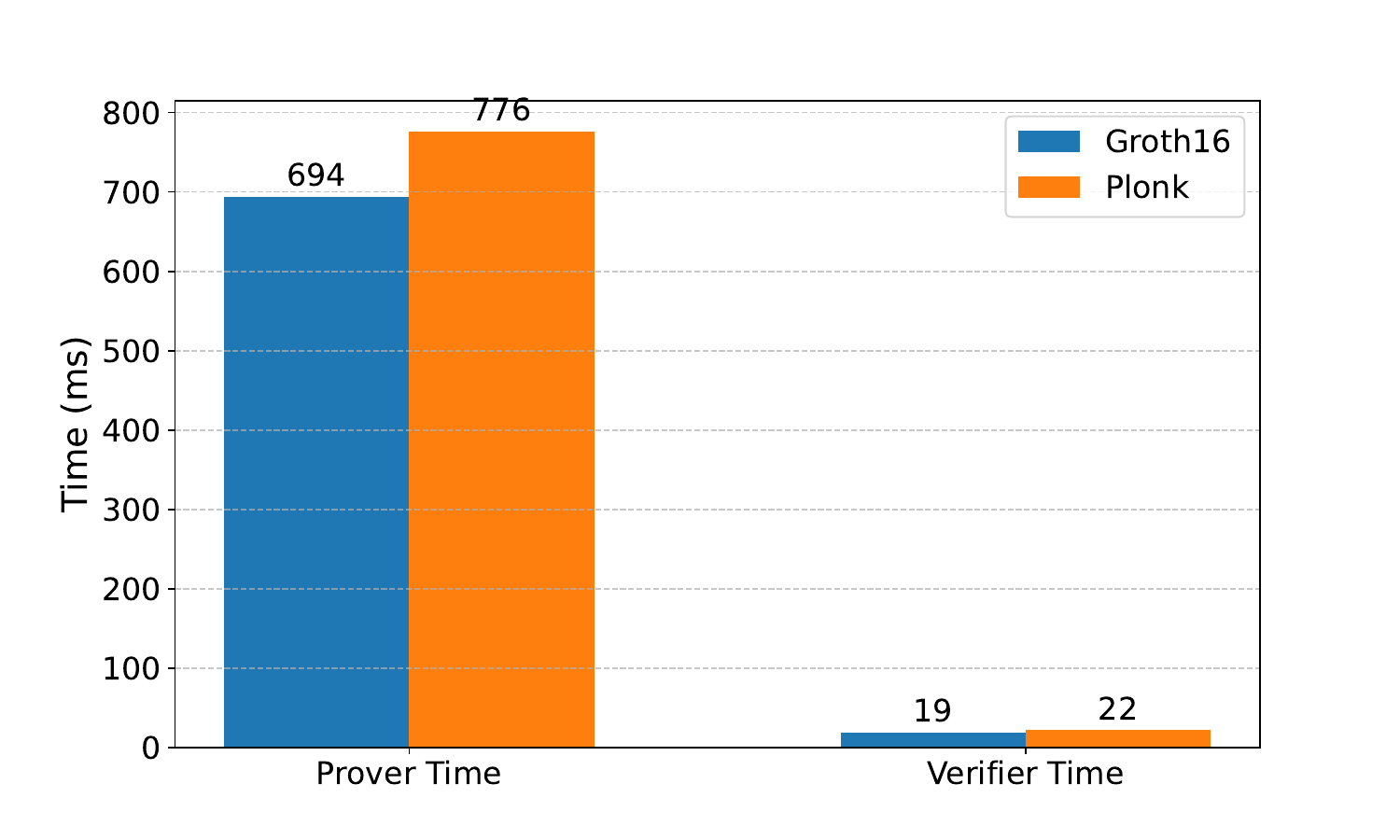}
\caption{Prover and verifier time: Groth16 vs. Plonk comparison.}
\label{fig:groth16plonk}
\end{figure}

 Figure \ref{fig:groth16plonk} presents the prover and verifier times for two distinct ZKP proving systems: Groth16 and Plonk. It is shown that Groth16 requires 694 milliseconds, which is less than Plonk's 777 milliseconds for prover time. Additionally, the verification times for Groth16 and Plonk are 19 milliseconds and 22 milliseconds, respectively. While Groth16 demonstrates faster processing times, it necessitates a circuit-specific ZKP key generation. In contrast, Plonk benefits from a universal setup, allowing the keys generated to be applied across various circuits.
\vspace{-3pt}

\par
\begin{figure}[h]
\centering
\includegraphics[width=0.47\textwidth]{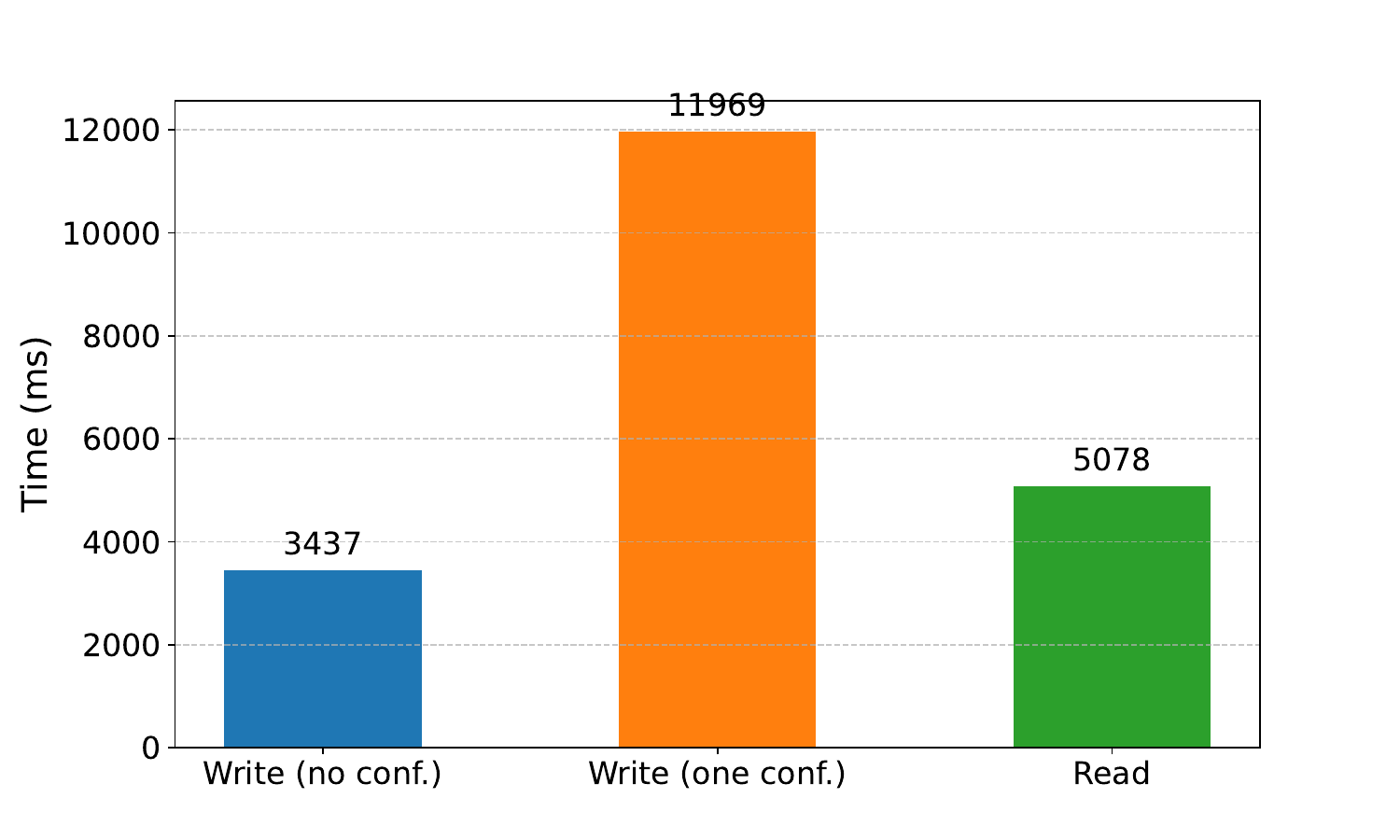}
\caption{Blockchain proof write and read time.}
\label{fig:blockchanproof}
\end{figure}

 Fig. \ref{fig:blockchanproof} illustrates the time required for writing and reading proofs on the blockchain within the zk-IoT framework. Specifically, submitting the proof to the smart contract for storage takes 3,437 milliseconds, writing the proof and receiving the first confirmation takes 11,969 milliseconds, and reading the proof requires 5,078 milliseconds. These durations highlight that blockchain operations are considerably slower than proof generation and verification, underscoring the need for faster blockchain technology that is more compatible with the zk-IoT framework.
\vspace{-2pt}
\section{Conclusion}
In this paper, we have introduced the zk-IoT framework as a groundbreaking development in enhancing the security of IoT ecosystems, utilizing ZKP and blockchain technology. This innovative framework is adept at verifying the correctness of IoT firmware execution and ensuring data integrity, even in scenarios where devices may be compromised. Our approach not only fortifies the security of IoT devices communications but also fosters new opportunities for dependable and autonomous interactions within IoT networks. These advancements are vital for the progress of smart city infrastructures, healthcare systems, and the realm of industrial automation. 

\par The practical implementation and preliminary performance analysis of the zk-IoT framework, as demonstrated in our work, confirm its practicality and effectiveness. This signals a shift towards a more secure and trustworthy era of IoT operations. We plan to present the detailed performance evaluation in a subsequent paper, following the completion of a full implementation.

\bibliographystyle{ieeetr}
\bibliography{output}

\end{document}